\documentclass[aps,prd,twocolumn,a4paper]{revtex4-1}
\usepackage{ulem}
\usepackage{color}
\usepackage{graphicx}
\usepackage{hyperref}   
\usepackage{appendix}
\usepackage{epsfig}
\usepackage{epstopdf}
\usepackage{amsmath}
\usepackage{subfig}
\usepackage{comment}
\usepackage[dvipsnames]{xcolor}
\newcommand{\be}{\begin{equation}}
\newcommand{\ee}{\end{equation}}
\newcommand{\ba}{\begin{eqnarray}}
\newcommand{\ea}{\end{eqnarray}}

\begin{document}

\title{Transport coefficients of hot magnetized QCD matter beyond the lowest Landau level approximation}
\author{Manu Kurian $^{a}$}
\email{manu.kurian@iitgn.ac.in}
\author{Sukanya Mitra $^{b}$}
\email{sukanya.mitra10@gmail.com}
\author{Snigdha Ghosh $^{a}$}
\email{snigdha.physics@gmail.com}
\author{Vinod Chandra $^{a}$}
\email{vchandra@iitgn.ac.in}
\affiliation{$^a$Indian Institute of Technology Gandhinagar, Gandhinagar-382355, Gujarat, India}
\affiliation{$^b$ National  Superconducting Cyclotron Laboratory, Michigan State University, East Lansing, Michigan 48824, USA}

\begin{abstract}
In this article, shear viscosity, bulk viscosity, and thermal conductivity of a hot QCD medium have been studied in the presence of strong magnetic field.  To model the hot magnetized QCD matter, an extended quasi-particle description of the hot QCD equation of state in the presence of the magnetic field has been adopted. The effects of higher Landau levels on the temperature dependence of viscous coefficients (bulk and shear viscosities) and thermal conductivity have been obtained by considering the $1\rightarrow 2$ processes in the presence of the strong magnetic field. An effective covariant kinetic theory has been set up in (1+1)-dimensional that includes mean field contributions in terms of quasi-particle dispersions and magnetic field to describe the Landau level dynamics of quarks. The sensitivity of these parameters to the magnitude of the magnetic field has also been explored. Both the magnetic field and mean field contributions have seen to play a significant role in obtaining the temperature behaviour of the transport coefficients of hot QCD medium.
\end{abstract}

\maketitle
 
\section{Introduction}
Relativistic heavy-ion collision (RHIC) experiments have 
reported the presence of strongly coupled matter- Quark-gluon 
plasma (QGP) as a near-ideal fluid~\cite{STAR,Aamodt:2010pb}. 
The quantitative estimation of the experimental observables such as  
the collective flow and transverse momentum spectra of the 
produced particles from the hydrodynamic simulations involve 
the dependence upon the transport parameters of the medium. 
Thus,  the transport coefficients  are the essential input parameters 
for the hydrodynamic evolution of the system.  

 Recent investigations show that intense magnetic field 
is created in the early stages of the non-central asymmetric 
collisions~\cite{Skokov:2009qp,Zhong:2014cda,deng,Das:2016cwd}. 
This magnetic field affects the thermodynamic and transport 
properties of the hot dense QCD matter produced in the RHIC.  
Ref~\cite{Inghirami:2016iru} describes the extension of 
ECHO-QGP~\cite{DelZanna:2013eua, Becattini:2015ska} to the 
magnetohydrodynamic regime. 
The recent major developments regarding the intense magnetic field 
in heavy-ion collision  include the chiral magnetic 
effect~\cite{Fukushima:2008xe,Sadofyev:2010pr, Huang}, 
chiral vortical effects~\cite{Kharzeev:2015znc,
Avkhadiev:2017fxj,Yamamoto:2017uul} 
and very recent realization  of global 
$\Lambda$-hyperon polarization 
in non-central RHIC~\cite{STAR:2017ckg,Becattini:2016gvu}.
This sets the motivation to study the transport coefficients in presence of the 
strong magnetic field.  The transport parameters under investigation are the 
viscous coefficients (shear and bulk) and the thermal conductivity of 
the hot magnetized QGP. Importance of the transport 
processes in RHIC is well studied~\cite{Luzum:2008cw} 
and reconfirmed by the recent ALICE 
results~\cite{Adam:2016izf,Abelev:2012pa,Abelev:2012pp}. 

 Quantizing quark/antiquark field 
in the presence of strong magnetic field background gives the 
Landau levels as energy eigenvalues. The quark/antiquark 
degrees of freedom is governed by $(1+1)-$dimensional Landau 
level kinematics whereas gluonic degrees of freedom remain 
intact in the presence of magnetic  field~\cite{Hattori:2017qih, Kurian:2017yxj}. 
However, gluons can be indirectly affected by the magnetic 
field through the quark loops while defining the Debye mass of the system.

 Shear and bulk viscosities can be estimated from 
Green-Kubo formulation both in the presence and absence 
of magnetic field~\cite{Hattori:2017qih,
Kharzeev:2007wb,Moore:2008ws,Czajka:2017bod}. 
Lattice results for 
the shear and bulk viscosities to entropy ratio are also 
well investigated~\cite{Nakamura:2004sy,Astrakhantsev:2017nrs,Astrakhantsev:2018oue}. 
Viscous pressure tensor quantifies the energy-momentum 
dissipation with the space-time evolution and is 
characterized by seven viscous coefficients in 
the strong magnetic field~\cite{Tuchin:2011jw}. 
The seven viscous coefficients consist of two bulk 
viscosities (both transverse and longitudinal) 
and five shear viscosities. The present investigations 
are focused on the longitudinal component (along the 
direction of $\vec{B}$) of shear and bulk viscosities
since other components of viscosities are negligible 
in the strong magnetic field. Another key transport 
coefficient under investigation is the thermal conductivity 
of the QGP medium. The temperature dependence of 
thermal conductivity has been studied in the absence of 
magnetic field in the Ref.\cite{Marty:2013ita}. 
The  shear and bulk viscosities, electric and thermal conductivities and their 
relative significance have been studied in Ref. \cite{Mitra:2017sjo} within a quasiparticle description of 
interacting hot QCD equations of state.
The first step 
towards the estimation of transport coefficients from 
the effective kinetic theory is to include proper 
collision integral for the processes in the strong 
field. This can be done within the relaxation time 
approximation (RTA). Microscopic processes or interactions 
are the inputs of the transport coefficients and are 
incorporated through thermal relaxation times. Note 
that the $1\rightarrow 2$ processes such as quark-antiquark pair production/annihilation 
are dominant in the presence 
of strong magnetic field~\cite{Fukushima:2017lvb, Hattori:2016lqx}.

 The prime focus of the present article is to estimate 
the temperature behaviour of the transport coefficients 
such as bulk viscosity, shear viscosity and thermal 
conductivity, incorporating the hot QCD medium effects 
in the presence of the strong magnetic field. Estimation 
of the transport parameters can be done in two 
equivalent approaches $viz.,$ the hard thermal 
loop effective theory (HTL)~\cite{Arnold:2003zc,
ValleBasagoiti:2002ir,Moore:2001fga} and the 
relativistic semi-classical transport
theory~\cite{Fukushima:2017lvb,Chen:2009sm,Khvorostukhin:2010cw,
Xu:2007ns,Thakur:2017hfc}. The present analysis 
is done with the relativistic transport theory by 
employing the Chapman-Enskog method. Hot QCD medium 
effects are encoded in the quark/antiquark and gluonic 
degrees of freedom by adopting the effective 
fugacity quasiparticle model (EQPM)~\cite{Kurian:2017yxj,Chandra:2011en,
Chandra:2007ca,Mitra:2016zdw}. The transport 
coefficients pick up the mean field term (force term) 
as described in Ref~\cite{Mitra:2018akk}. The mean field term comes from 
the local conservations of number current and stress-energy 
tensor in the covariant effective kinetic theory. 
In the current analysis, we 
investigate the mean field 
corrections in the presence of strong 
magnetic field and study the temperature behaviour of 
the transport coefficients. 
Here, the strong magnetic field 
restricts the calculations to $(1+1)-$dimensional 
(dimensional reduction) covariant effective kinetic theory for quarks and antiquarks.

 The manuscript is organized as follows. In section II, 
the mathematical formulation for the estimation of 
transport coefficients from the effective covariant 
kinetic theory is discussed along with the quasiparticle 
description of hot QCD medium in the strong magnetic field. 
Section III deals with the thermal relaxation for the 
$1\rightarrow2$ processes in the strong magnetic field. Predictions of the 
transport coefficients in the magnetic field are 
discussed in section IV.  Finally, in section V the 
summary and outlook of the are presented.

\section{Formalism: Transport coefficients at strong magnetic field}
The strong magnetic field $\vec{B}=B\hat{z}$ constraints 
the quarks/antiquarks motion parallel to field with 
a transverse density of states. 
The viscous coefficients~\cite{Hattori:2017qih,Kurian:2018dbn} and heavy quark diffusion 
coefficient~\cite{Fukushima:2015wck} have been 
perturbatively calculated under the regime $\alpha_{s}\mid q_feB\mid
\ll T^{2}\ll \mid q_feB\mid$ with the lowest Landau level (LLL) approximation.
But the validity of LLL approximation is questionable since higher Landau level 
contributions are significant at $\mid eB\mid=10 m_{\pi}^2$ in the temperature range above $200$ MeV. 
Here, we are focusing on the more realistic regime $gT
\ll \sqrt{\mid q_feB\mid}$ in which higher Landau level (HLL) contributions are significant. In the 
very recent work~\cite{Fukushima:2017lvb}, Fukushima and Hidaka have been estimated 
the longitudinal conductivity of 
magnetized QGP with full Landau level resummation in the regime $gT
\ll \sqrt{\mid q_feB\mid}$.

 The formalism for the estimation of transport coefficients 
includes the quasiparticle modeling of the system 
away from the equilibrium followed by the 
setting up of the effective kinetic theory for 
different processes. Quasiparticle models encode the medium effects, 
$viz.$, effective fugacity or with effective mass. 
The later include self-consistent and
single parameter quasiparticle models~\cite{Bannur:2006js}, 
NJL and PNJL based quasiparticle models~\cite{Dumitru}, 
effective mass with Polyakov loop~\cite{D'Elia:97} 
and recently proposed quasiparticle models based on the
Gribov-Zwanziger (GZ) quantization~\cite{Su:2014rma,
zwig,Bandyopadhyay:2015wua}. Here, the analysis is done within the effective fugacity 
quasiparticle model (EQPM) where the medium interactions are encoded through 
temperature dependent  effective quasigluon and quasiquark/antiquark fugacities,  
$z_{g}$ and $z_{q}$ respectively. The extended EQPM describes the hot QCD medium effects 
in strong magnetic field~\cite{Kurian:2017yxj}. 
We considered the (2+1) flavor lattice QCD equation of state (EoS) 
(LEoS)~\cite{Cheng:2007jq,Borsanyi} and the 3-loop HTLpt 
EOS~\cite{Haque,Andersen} for the effective description of 
QGP in strong magnetic field~\cite{Kurian:2017yxj,Kurian:2018dbn}.

\subsection*{Transport coefficients from effective (1+1)-D kinetic theory}
In the absence of magnetic field, the particle four flow $\bar{N}^{\mu}(x)$ can be defined 
 in terms of quasiparticle (dressed) momenta $\vec{\bar{p}}_k$ 
 within EQPM as~\cite{Mitra:2018akk},
\begin{align}\label{11}
\bar{N}^{\mu}(x)&=\sum_{k=1}^{N}\nu_k\int{\dfrac{d^3\mid
\vec{\bar{p}}_k\mid}{(2\pi)^3{\omega}_k}
\bar{p}_k^{\mu}f^0_k(x,\bar{p}_k)}\nonumber\\
&+\sum_{k=1}^{N}\delta\omega\nu_k\int{\dfrac{d^3\mid
\vec{\bar{p}}_k\mid}{(2\pi)^3{\omega}_k}
\dfrac{\langle\bar{p}_k^{\mu}\rangle}{E_{k}} f^0_k(x,\bar{p}_k)},
\end{align}
in which $\nu_k$ is the degeneracy factor of the $k^{th}$ 
species. Here, we are considering non-zero masses ($m_f$) for quarks (up, down and strange quarks with 
masses $m_u=3$ MeV, $m_d=5$ MeV and $m_s=100$ MeV respectively) and hence $E_{k}=\sqrt{
{\mid\vec{\bar{p}}_k\mid}^2+m_f^2}$ for quarks/antiquarks and 
$E_{k}=\mid\vec{\bar{p}}\mid$ for gluons.  
The term $\langle\bar{p}^{\mu}\rangle=\Delta^{\mu\nu}
\bar{p}_{\nu}$ is the irreducible tensor with 
$\Delta^{\mu\nu}=g^{\mu\nu}-u^{\mu}u^{\nu}$ as 
the projection operator. The metric has the 
form $g^{\mu\nu}=$diag $(1,-1,-1,-1)$. The quasiquark distribution 
function in local rest frame with the hydrodynamic four-velocity  
$u^{\mu}\equiv (1,\bf{0})$ is given by, 
\begin{equation}\label{4}
f^0_{q, g}=\dfrac{z_{q, g}\exp{[-\beta (u^{\mu}p_{\mu})]}}{1\pm z_{q, g}\exp{[-\beta (u^{\mu}p_{\mu})]}},
\end{equation}
 with $p^{\mu}=(E, \vec{\bar{p}})$. 
 Quasiparticle momenta (dressed momenta)
 and bare particle four-momenta 
can be related from the dispersion relations as,
\begin{align}\label{7}
\bar{p}^{\mu}&=p^{\mu}+\delta\omega u^{\mu},
& \delta\omega= T^{2}\partial_{T} \ln(z_{q, g}),
\end{align}
which modifies the zeroth component of the 
four-momenta in the local rest frame. Hence, we have
\begin{equation}\label{7.0}
\bar{p}^{0}\equiv\omega_{k}=E_{k}+\delta\omega.
\end{equation} 
The dispersion relation in Eq.~(\ref{7.0})
encodes the collective excitation of 
quasiparton along with the single particle energy. 
Also, the energy-momentum tensor $\bar{T}^{\mu\nu}$ in terms of dressed momenta
takes the following form,
\begin{align}\label{7.2}
\bar{T}^{\mu\nu}(x)&=\sum_{k=1}^{N}
\nu_k\int{\dfrac{d^3\mid
\vec{\bar{p}}_k\mid}{(2\pi)^3\omega_{k}}
\bar{p}_k^{\mu}\bar{p}_k^{\nu}f^0_k(x,\bar{p}_{k})}\nonumber\\
&+\sum_{k=1}^{N}\delta\omega
\nu_k\int{\dfrac{d^3\mid
\vec{\bar{p}}_k\mid}{(2\pi)^3
{\omega}_k}\dfrac{\langle\bar{p}_k^{\mu}\bar{p}_k^{\nu}\rangle}
{E_k} f^0_k(x,\bar{p}_{k})},
\end{align}
where $\langle\bar{p}_k^{\mu}\bar{p}_k^{\nu}\rangle=
\dfrac{1}{2}(\Delta^{\mu\alpha}\Delta^{\nu\beta}+
\Delta^{\mu\beta}\Delta^{\nu\alpha})\bar{p}_{\alpha}\bar{p}_{\beta}$.

 In our case, Eq.~(\ref{7.2}) should 
 rewritten for the hot QCD medium in the strong magnetic field $\vec{B}=B\hat{z}$ limit. 
 Thereafter, the transport coefficients could be obtained by realizing the 
 microscopic (transport theory) definition of $\bar{T}^{\mu\nu}$ to the macroscopic decomposition 
 at various order.  Recall that the EQPM in the 
 presence of a strong magnetic field is studied by considering the 
 Landau level dynamics in the dispersion relation for quarks 
 whereas gluonic part remain invariant in magnetic field~\cite{Kurian:2017yxj,Kurian:2018dbn}. 
 The quasi-quark/antiquark distribution function in the strong magnetic field 
 background takes the form as in Eq.~(\ref{4}) with the particle four-momenta 
 $p_{\|}^{\mu}=(\omega_l,0,0,\bar{p}_z)$. The zeroth component of four-momenta becomes,
 \begin{equation}\label{7.1}
\bar{p}^{0}\equiv\omega_{l}=\sqrt{\bar{p}_{z}^{2}+m_f^{2}+2l\mid q_feB\mid}+\delta\omega.
\end{equation} 
where $\sqrt{\bar{p}_{z}^{2}+m_f^{2}+2l\mid q_feB\mid}\equiv E_{l}$ is the 
Landau level energy eigenvalue in the strong magnetic field. 

 Macroscopically, the energy-momentum tensor in the presence 
of magnetic field $\vec{B}=B\hat{z}$ 
can be decomposed as~\cite{Hattori:2017qih},
\begin{equation}\label{a1}
\bar{T}^{\mu\nu}=\varepsilon u^{\mu}u^{\nu}-P_{\bot}\Xi^{\mu\nu}+P_{\|}b^{\mu}b^{\nu}+\tau^{\mu\nu},
\end{equation}
where $u^{\mu}$ is the flow vector and $b^{\mu}=\epsilon^{\mu
\nu\alpha\beta}F_{\nu\alpha}u_{\beta}/(2B)$ 
with $B=\sqrt{-B^{\mu}B_{\mu}}$. Here, $P_{\bot}$ and $P_{\|}$ are the transverse 
and longitudinal components of pressure respectively and holds the relation 
$P_{\bot}=P_{\|}-MB$, where the magnetization $M={(\frac{\partial P}{\partial B})}_{T}$. 
The tensor $\Xi^{\mu\nu}=\Delta^{\mu\nu}+b^{\mu}b^{\nu}$, projects out the 
two-dimensional space orthogonal 
to both the flow and magnetic field. 
In the presence of strong magnetic field, the pressure can be defined as, 
\begin{equation}
P={P_{\|}}_q+P_g,
\end{equation}
with $P\equiv \bar{T}^{\mu\nu}b_{\mu}b_{\nu}=\bar{T}^{33}$.
Here, ${P_{\|}}_q$ is the dominant quark and antiquark contribution to the pressure  
in the strong magnetic field~\cite{Kurian:2017yxj,Hattori:2017qih} and have the following form,
\begin{equation}\label{a2}
P_{\|}=\sum_{l=0}^{\infty}\dfrac{\mid q_{f}eB\mid }{\pi^2}N_c
 \int_{0}^{\infty}{dp_z\dfrac{p_z^2}{E_{l}}\mu_lf^0_q}.
\end{equation} 
The integration phase factor in the strong field 
due to dimensional reduction~\cite{Bruckmann:2017pft,Tawfik:2015apa,
Gusynin:1995nb} is defined as, 
\begin{equation}\label{13}
\int{\dfrac{d^{3}p}{(2\pi)^{3}}}\rightarrow
\dfrac{\mid q_feB\mid}{2\pi}\int_{-\infty}^{\infty}{\dfrac{dp_{z}}{2\pi}\mu_l},
\end{equation} 
where $\mu_l=(2-\delta_{l0})$ is the spin degeneracy factor of the Landau levels. 
Since gluonic dynamics are not directly affected by the magnetic field, the gluonic contribution 
$P_g$ retains the same form as in the absence of 
magnetic field and is well investigated in the work~\cite{Chandra:2011en}.
Note that in the presence of the strong magnetic field quark/antiquark contribution is 
dominant compared with that of gluons~\cite{Hattori:2016lqx, Fukushima:2017lvb, Hattori:2017qih}.
 Also, we can define 
the quark and antiquark contribution to energy density in the strong field as, 
\begin{equation}\label{a3}
 \varepsilon_{\|}=\sum_{l=0}^{\infty}\dfrac{\mid q_{f}eB\mid }{\pi^2}N_c
 \int_{0}^{\infty}{dp_z\dfrac{{(\omega_p)}^2}{\omega_l}\mu_lf^0_q}.
\end{equation}
Since the quark dynamics is constrained in the $(1+1)$-dimensional space, 
both $b^{\mu}$ and $u^{\mu}$ are 
longitudinal $(1+1)$-dimensional vector and at the same time $b^{\mu}$ 
is orthogonal to $u^{\mu}$. The longitudinal projection operator 
$\Delta_{\|}^{\mu\nu}$ is perpendicular to $u^{\mu}$ and can 
constructed from $b^{\mu}$~\cite{Li:2017tgi} as,
\begin{equation}\label{a4}
\Delta_{\|}^{\mu\nu}\equiv g_{\|}^{\mu\nu}-u^{\mu}u^{\nu}=- b^{\mu}b^{\nu},
\end{equation}
where $g_{\|}^{\mu\nu}=$ diag $(1,0,0,-1)$. Hence, in the strong magnetic field, 
the equilibrium energy-momentum tensor from the quark/antiquark part takes the form as follows,
\begin{equation}\label{a6}
T^{\mu\nu}=\varepsilon_{\|} u^{\mu}u^{\nu}-P_{\|}\Delta_{\|}^{\mu\nu}.
\end{equation}
In the strong magnetic field, $T^{\mu\nu}$ can be defined in terms 
of quasiparticle momenta of quarks and antiquarks as the following,
\begin{align}\label{14}
T^{\mu\nu}(x)&=\sum_{l=0}^{\infty}\sum_{k\in q,\bar{q}}\mu_l\dfrac{\mid {q_f}_keB\mid}{2\pi}
N_c\int_{-\infty}^{\infty}{\dfrac{d\bar{p}_{z_k}}{(2\pi)\omega_{l_k}}
\bar{p_{\|}}_k^{\mu}\bar{p_{\|}}_k^{\nu}}\nonumber\\
&\times f^0_k(x,\bar{p}_{z_k})\nonumber\\
&+\sum_{l=0}^{\infty}\sum_{k\in q,\bar{q}}\delta\omega\mu_l\dfrac{\mid {q_f}_keB\mid}{2\pi}
N_c\int_{-\infty}^{\infty}{\dfrac{d\bar{p}_{z_k}}{(2\pi)
{\omega_{l_k}}}\dfrac{\langle\bar{p_{\|}}_k^{\mu}\bar{p_{\|}}_k^{\nu}\rangle}
{E_{l_k}} }\nonumber\\
&\times f^0_k(x,\bar{p}_{z_k}),
\end{align}
which give back the expressions as in Eqs.~(\ref{a2}) and~(\ref{a3}) 
for the pressure and energy density respectively through the following definitions,
\begin{align}\label{7}
\varepsilon_{\|}&=u^{\mu}u^{\nu}T_{\mu\nu},
& P_{\|}= \Delta_{\|}^{\mu\nu}T_{\mu\nu}.
\end{align}
Here, $\bar{p_{\|}}_k^{\mu}\equiv (\omega_{l_k},0,0,p_{z_k})$ 
incorporates the longitudinal components and $\langle\bar{p_{\|}}_k^{\mu}\bar{p_{\|}}_k^{\nu}\rangle=
\dfrac{1}{2}(\Delta_{\|}^{\mu\alpha}\Delta_{\|}^{\nu\beta}+
\Delta_{\|}^{\mu\beta}\Delta_{\|}^{\nu\alpha})\bar{p_{\|}}_{\alpha}\bar{p_{\|}}_{\beta}$.  
For the weak (moderate) magnetic field, one also needs to analyse the transverse 
dynamics of the hot QCD matter. In these situations, the transverse components of 
various transport coefficients might play a significant role. These aspects are beyond 
the scope of the present work and the matter of future extensions of the work.
Following the above arguments, four flow $N^{\mu}$ of the quarks and antiquarks in the strong magnetic field has 
the following form,
\begin{align}\label{12}
N^{\mu}(x)&=\sum_{l=0}^{\infty}\sum_{k\in q,\bar{q}}\mu_l\dfrac{\mid {q_f}_keB\mid}{2\pi}
N_c\int_{-\infty}^{\infty}{\dfrac{d\bar{p}_{z_k}}{(2\pi)\omega_{l_k}}
\bar{p_{\|}}_k^{\mu}}\nonumber\\
&\times f^0_k(x,\bar{p}_{z_k})\nonumber\\
&+\sum_{l=0}^{\infty}\sum_{k\in q,\bar{q}}\delta\omega\mu_l\dfrac{\mid {q_f}_keB\mid}
{2\pi}N_c\int_{-\infty}^{\infty}{\dfrac{d\bar{p}_{z_k}
}{(2\pi){\omega_{l_k}}}\dfrac{\langle\bar{p_{\|}}_k^{\mu}\rangle}
{E_{l_k}}}\nonumber\\
&\times f^0_k(x,\bar{p}_k),
\end{align}
with $\langle\bar{p_{\|}}^{\mu}\rangle={\Delta_{\|}}^{\mu\nu}
\bar{p_{\|}}_{\nu}$.

  Estimation of the transport coefficients requires the 
system away from equilibrium. 
In the current analysis, we are focusing 
on the dominant quark/antiquark dynamics of the magnetized QGP.  Here, we need to set-up the 
relativistic transport equation, which quantifies the 
rate of change of quasiquark/antiquark distribution function in terms of collision 
integral. The thermal relaxation time ($\tau_{\text{eff}}$) linearize the collision 
term ($C(f_{q})$) in the following way,
\begin{equation}\label{15}
\dfrac{1}{\omega_{l_{k}}}\bar{p_{\|}}^{\mu}_k\partial_{\mu}f^0_k(x,\bar{p}_{z_k})+F_z\dfrac{\partial f_k^0}{\partial p_{z_k}}=C(f_{k})\equiv
-\dfrac{\delta f_k}{\tau_{\text{eff}}},
\end{equation}
with $F_z=-\partial_{\mu}(\delta\omega u^{\mu}u_z)$ is 
the force term from the conservation of particle density 
and energy momentum~\cite{Mitra:2018akk}.
The local momentum distribution function 
of quarks can expand as, 
\begin{align}
f_k&= f^{0}_k(p_z)+\delta f_k,    &\delta f_k=f_k^0(1\pm f_k^0)\phi_k.
\end{align}
Here, $\phi_k$ defines the deviation of the quasiquark distribution 
function from its equilibrium. The Eq.~(\ref{15}) gives 
the effective kinetic theory 
description of the quasipartons under EQPM in the strong 
magnetic field. In order to estimate the transport 
coefficients, we employ the Chapman-Enskog (CE) method.
Applying the definition of equilibrium quasiparton 
momentum distribution function as in Eq.~(\ref{4}), 
the first term of Eq.~(\ref{15}) gives the number of 
terms with thermodynamic forces of the transport processes. 
The second term of Eq.~(\ref{15}) vanishes for a 
co-moving frame. Finally, we are left with,
\begin{equation}\label{16}
Q_kX+\langle\bar{p_{\|}}_k^{\mu}\rangle(\omega_{l_k}-h_k)
X_{q\mu}-\langle\langle\bar{p_{\|}}_k^{\mu}\bar{p_{\|}}_k^{\nu}
\rangle\rangle X_{\mu\nu}=-\dfrac{T\omega_{p_k}}{\tau_{\text{eff}}}\phi_k,
\end{equation}
in which the conformal factor due to the dimensional 
reduction in the strong field limit is $Q_k=(\bar{p}^2_{z_k}
-\omega_{l_k}^2c^2_s)$ where $c^2_s$ is the speed of sound 
and $h_k$ is the enthalpy per particle of the system that 
can be defined from the basic QCD thermodynamics. 
Here, $\langle\langle P_{\|}^{\mu}R_{\|}^{\nu}\rangle\rangle
=\left\{ \frac{1}{2}{\Delta_{\|}}^{\mu}_{\alpha}{\Delta_{\|}}^{\nu}_{\beta}
+\frac{1}{2}{\Delta_{\|}}^{\mu}_{\beta}{\Delta_{\|}}^{\nu}_{\alpha}
-\frac{1}{3}{\Delta_{\|}}_{\alpha\beta}\Delta_{\|}^{\mu\nu}
\right\} P_{\|}^{\alpha}R_{\|}^{\beta}$. The bulk viscous force, thermal force and shear 
viscous force are defined respectively as follows,
\begin{align}\label{17}
&X=\partial.u,\\
&X_q^{\mu}=\left\{\dfrac{\bigtriangledown^{\mu}T}
{T}-\dfrac{\bigtriangledown^{\mu}P}{nh}\right\},\\
&X_{\mu\nu}=\langle\langle\partial_{\mu}u_{\nu}\rangle\rangle,
\end{align}
where $h$ is the total enthalpy defined as $h=\sum_{k=0}^{N}h_k$ and 
$n$ is the total number density of the system.
Note that here $\mu=0,3$ describes only the 
longitudinal components in the strong magnetic 
field. Also, the deviation function $\phi_k$ 
that is the linear combination of these forces 
can be represented as,
\begin{equation}\label{18}
\phi_{k}=A_kX+B_k^{\mu}X_{q\mu}-C^{\mu\nu}_kX_{\mu\nu},
\end{equation}
where the coefficients can be defined from Eq.~(\ref{16}) as,
\begin{align}\label{19}
&A_k=\dfrac{Q_k}{\lbrace-\frac{T\omega_{l_k}}
{\tau_{\text{eff}}}\rbrace},\\
&B^{\mu}_k=\langle\bar{p}_k^{\mu}\rangle\dfrac{(\omega_{l_k}-h_k)}
{\lbrace-\frac{T\omega_{l_k}}{\tau_{\text{eff}}}\rbrace},\\
&C_k^{\mu\nu}=\dfrac{\langle\langle\bar{p_{\|}}_k^{\mu}\bar{p_{\|}}_k^{\nu}\rangle\rangle}
{{\lbrace-\frac{T\omega_{l_k}}{\tau_{\text{eff}}}\rbrace}}.
\end{align}
Following this formalism, we can estimate the 
viscous coefficients and thermal conductivity 
of the QGP medium in the strong magnetic field.
\subsubsection{Shear and bulk viscosity}
We can define the  pressure tensor from the 
energy-momentum tensor as in the following way,
\begin{equation}\label{20}
P^{\mu\nu}={\Delta_{\|}}_{\sigma}^{\mu}T^{\sigma\tau}
{\Delta_{\|}}_{\tau}^{\nu}.
\end{equation}
We can decompose the $P^{\mu\nu}$ in equilibrium 
and non-equilibrium components of distribution 
function as follows,
\begin{equation}\label{21}
P^{\mu\nu}=-P{\Delta_{\|}}^{\mu\nu}+\Pi^{\mu\nu},
\end{equation}
where $\Pi^{\mu\nu}$ is the viscous pressure tensor. 
Following the definition of $T^{\mu\nu}$ as in 
Eq.~(\ref{14}), $\Pi^{\mu\nu}$ takes the form,
\begin{align}\label{22}
\Pi^{\mu\nu}&=\sum_{l=0}^{\infty}\sum_{k\in q, \bar{q}}\mu_l\dfrac{\mid {q_f}_keB\mid}{2\pi}N_c
\int_{-\infty}^{\infty}{\dfrac{d\bar{p}_{z_k}}{(2\pi)\omega_{l_k}}
\langle\bar{p_{\|}}_k^{\mu}\bar{p_{\|}}_k^{\nu}\rangle}\nonumber\\
&\times \delta f_k(x,\bar{p}_{z_k})\nonumber\\
&+\sum_{l=0}^{\infty}\sum_{k\in q, \bar{q}}\delta\omega\mu_l\dfrac{\mid {q_f}_keB\mid}{2\pi}
N_c\int_{-\infty}^{\infty}{\dfrac{d\bar{p}_{z_k}}{(2\pi){\omega_{l_k}}}
\dfrac{\langle\bar{p_{\|}}_k^{\mu}\bar{p_{\|}}_k^{\nu}\rangle}
{E_{l_k}}}\nonumber\\
&\times \delta f_k(x,\bar{p}_{z_k}).
\end{align}
In the very strong magnetic field, the pressure 
tensor has different form as compared to the case 
without magnetic field. This is due to the $(1+1)-$dimensional 
energy eigenvalues of the quarks and antiquarks. 
Hence, $\mu$ and $\nu$ can be 0 or 3 in the strong 
magnetic field, describing the longitudinal components 
of the viscous pressure tensor. The form of viscous 
pressure tensor in the strong magnetic field is 
described in the recent works by Tuchin~\cite{Tuchin:2011jw,Tuchin:2013ie}. 
Magnetized plasma
is characterized by five shear components. Among the
five coefficients, four components are negligible
when the strength of the magnetic field is sufficiently
higher than the square of the temperature~\cite{Ofengeim:2015qxz}. 
Here, we are focusing on the non-negligible longitudinal 
component of shear and bulk viscous coefficients of the hot QGP medium 
in the strong magnetic field. 

Following~\cite{Mitra:2017sjo}, the longitudinal 
shear viscous tensor has the following form,
\begin{align}\label{23}
\bar{\Pi}^{\mu\nu}&=\Pi^{\mu\nu}-\Pi{\Delta_{\|}}^{\mu\nu}\nonumber\\
&=\sum_{l=0}^{\infty}\sum_{k\in q,\bar{q}}\mu_l\dfrac{\mid {q_f}_keB\mid}{2\pi}N_c\int_{-\infty}^{\infty}{\dfrac{d\bar{p}_{z_k}}
{(2\pi)\omega_{l_k}}\langle\langle\bar{p_{\|}}_k^{\mu}\bar{p_{\|}}_k^{\nu}
\rangle\rangle}\nonumber\\
&\times f^0_k(1-f^0_k)\phi_k\nonumber\\
&+\sum_{l=0}^{\infty}\sum_{k\in q,\bar{q}}\delta\omega\mu_l\dfrac{\mid {q_f}_keB\mid}{2\pi}N_c
\int_{-\infty}^{\infty}{\dfrac{d\bar{p}_{z_k}}{(2\pi){\omega_{l_k}}}
\dfrac{\langle\langle\bar{p_{\|}}_k^{\mu}\bar{p_{\|}}_k^{\nu}
\rangle\rangle}{E_{l_k}} }\nonumber\\
&\times f^0_k(1-f^0_k)\phi_k.
\end{align}
Also, the bulk viscous part in the longitudinal 
direction comes out to be,
\begin{align}\label{24}
\Pi&=\sum_{l=0}^{\infty}\sum_{k\in q,\bar{q}}\mu_l\dfrac{\mid {q_f}_keB\mid}{2\pi}\dfrac{N_c}{3}
\int_{-\infty}^{\infty}{\dfrac{d\bar{p}_{z_k}}{(2\pi)\omega_{l_k}}
{\Delta_{\|}}_{\mu\nu}}\nonumber\\
&\times\bar{p_{\|}}_k^{\mu}\bar{p_{\|}}_k^{\nu}
 f^0_k(1-f^0_k)\phi_k\nonumber\\
&+\sum_{l=0}^{\infty}\sum_{k\in q,\bar{q}}\delta\omega\mu_l\dfrac{\mid {q_f}_keB\mid}{2\pi}
\dfrac{N_c}{3}\int_{-\infty}^{\infty}{\dfrac{d\bar{p}_{z_k}}
{(2\pi){\omega_{l_k}}}{\Delta_{\|}}_{\mu\nu}}\nonumber\\
&\times\dfrac{\bar{p_{\|}}_k^{\mu}\bar{p_{\|}}_k^{\nu}}
{E_{l_k}}  f^0_k(1-f^0_k)\phi_k.
\end{align}
Substituting $\phi_k$ from Eq.~(\ref{18}) and 
comparing with the macroscopic definition 
$\Pi^{\mu\nu}=2\eta\langle\langle\partial^{\mu}u^{\nu}
\rangle\rangle+\zeta{\Delta_{\|}}^{\mu\nu}\partial.u$, we can 
obtain the expressions of longitudinal viscosity 
coefficients in the strong field limit. Note that 
the longitudinal component of shear viscosity, 
i.e., in the direction of magnetic field, is 
defined from $\bar{\Pi}^{33}$~\cite{Ofengeim:2015qxz}. The 
longitudinal shear $\eta$ and bulk viscosity 
$\zeta$ are obtained as,
\begin{align}\label{25}
\eta&=\sum_{l=0}^{\infty}\sum_{k\in q,\bar{q}}\mu_l\dfrac{\mid {q_f}_keB\mid}{\pi}\dfrac{N_c}{9T}
\int_{-\infty}^{\infty}{\dfrac{d\bar{p}_{z_k}}{(2\pi)}\dfrac{\mid
\bar{p}_{z_k}\mid^4}{\omega^2_{l_k}} \tau_{\text{eff}}
f^0_k(1-f^0_k)}\nonumber\\
&+\sum_{l=0}^{\infty}\sum_{k\in q,\bar{q}}\delta\omega\mu_l\dfrac{\mid {q_f}_keB\mid}{\pi}
\dfrac{N_c}{9T}\int_{-\infty}^{\infty}{\dfrac{d\bar{p}_{z_k}}
{(2\pi)}\dfrac{1}{\omega^2_{l_k}}\dfrac{\mid\bar{p}_{z_k}\mid^4}
{E_{l_k}}}\tau_{\text{eff}}\nonumber\\
&\times f^0_k(1-f^0_k),
\end{align}
and
\begin{align}\label{26}
\zeta&=\sum_{l=0}^{\infty}\sum_{k\in q,\bar{q}}\mu_l\dfrac{\mid {q_f}_keB\mid}{2\pi}\dfrac{N_c}
{3T}\int_{-\infty}^{\infty}{\dfrac{d\bar{p}_{z_k}}{(2\pi)}\dfrac{1}
{\omega^2_{l_k}}\lbrace \bar{p}^2_{z_k}
-\omega^2_{l_k}c_s^2} \rbrace^2\nonumber\\
&\times \tau_{\text{eff}}
f^0_k(1-f^0_k)\nonumber\\
&+\sum_{l=0}^{\infty}\sum_{k\in q,\bar{q}}\delta\omega\mu_l\dfrac{\mid {q_f}_keB\mid}{2\pi}
\dfrac{N_c}{3T}\int_{-\infty}^{\infty}{\dfrac{d\bar{p}_{z_k}}
{(2\pi)}\dfrac{1}{\omega^2_{l_k}}}\lbrace \bar{p}^2_{z_k}
-\omega^2_{l_k}c_s^2\rbrace^2\nonumber\\
&\times\dfrac{1}{E_{l_k}}\tau_{\text{eff}}f^0_k(1-f^0_k).
\end{align}
The second term in the Eq.~(\ref{25}) and 
Eq.~(\ref{26}) gives correction to viscous 
coefficients due to the quasiparton excitations 
whereas the first term comes from the usual 
kinetic theory of bare particles.
\subsubsection{Thermal conductivity}
The heat flow is the difference between the 
energy flow and enthalpy flow by the particle,
\begin{equation}\label{27}
I^{\mu}_q=u_{\nu}T^{\nu\sigma}{\Delta_{\|}}_{\sigma}^{\mu}
-hN^{\sigma}{\Delta_{\|}}_{\sigma}^{\mu}.
\end{equation}
In terms of the modified/non-equilibrium distribution function Eq.~(\ref{27}) becomes,
\begin{align}\label{28}
&I^{\mu}=u_{\nu}{\Delta_{\|}}^{\mu}_{\sigma}
\sum_{l=0}^{\infty}\sum_{k\in q,\bar{q}}\mu_l\dfrac{\mid {q_f}_keB\mid}{2\pi}N_c\int_{-\infty}^{\infty}{\dfrac{d\bar{p}_{z_k}}
{(2\pi)\omega_{l_k}}\bar{p_{\|}}_k^{\nu}\bar{p_{\|}}_k^{\sigma}}\delta f_k\nonumber\\
&-h{\Delta_{\|}}^{\mu}_{\sigma}\Bigg[\sum_{l=0}^{\infty}\sum_{k\in q,\bar{q}}\mu_l\dfrac{\mid {q_f}_keB\mid}
{2\pi}N_c\int_{-\infty}^{\infty}{\dfrac{d\bar{p}_{z_k}}{(2\pi)\omega_{l_k}}
\bar{p_{\|}}_k^{\sigma}\delta f_k(x,\bar{p}_{z_k})}\nonumber\\
&+\sum_{l=0}^{\infty}\sum_{k\in q,\bar{q}}\delta\omega\mu_l\dfrac{\mid {q_f}_keB\mid}{2\pi}N_c
\int_{-\infty}^{\infty}{\dfrac{d\bar{p}_{z_k}}{(2\pi){\omega_{l_k}}}
\dfrac{\langle\bar{p_{\|}}_k^{\sigma}\rangle}{E_{l_k}}}
 \delta f_k(x,\bar{p}_{z_k})\Bigg],
\end{align}
in which heat flow retains only non-equilibrium part 
of the distribution function. After contracting with 
projection operator and hydrodynamic velocity along 
with the substitution of $\delta f_k$ from Eq.~(\ref{15}) 
and comparing with the macroscopic definition of heat flow, 
we obtain
\begin{equation}\label{29}
I^{\mu}=\lambda TX_q^{\mu}.
\end{equation}
We obtain the thermal conductivity in the 
strong magnetic field as,
\begin{align}\label{30}
\lambda&= \Bigg\{\sum_{l=0}^{\infty}\sum_{k\in q,\bar{q}}\mu_l\dfrac{\mid {q_f}_keB\mid}{2\pi}
\dfrac{N_c}{T^2}\int_{-\infty}^{\infty}{\dfrac{d\bar{p}_{z_k}}
{(2\pi)}\tau_{\text{eff}}\dfrac{(\omega_{l_k}-h_k)^2}
{\omega^2_{l_k}}}\nonumber\\
&\times\mid\bar{p}_{z_k}\mid^2
f^0_k(1-f^0_k) \Bigg\}\nonumber\\
&- \Bigg\{\sum_{l=0}^{\infty}\sum_{k\in q,\bar{q}}\mu_l\delta\omega\dfrac{\mid {q_f}_keB\mid}
{2\pi}\dfrac{N_c}{T^2}\int_{-\infty}^{\infty}{\dfrac{d\bar{p}_{z_k}}
{(2\pi)}\tau_{\text{eff}}\dfrac{h_k(\omega_{l_k}-h_k)}
{\omega^2_{l_k}}}\nonumber\\
&\times\dfrac{\mid\bar{p}_{z_k}\mid ^2}{E_{l_k}}
f^0_k(1-f^0_k) \Bigg\}.
\end{align}
The second term with $\delta\omega$ in the heat 
flow comes from the $N_{\mu}$ which encodes the 
quasiparticle excitation in the thermal conductivity.

\section{Thermal relaxation in the strong magnetic field}
Thermal relaxation is the essential dynamical 
input of the transport processes which counts 
for the microscopic interaction of the system. 
In the strong magnetic field, the 
1 $\rightarrow$ 2 processes (gluon to quark-antiquark pair) 
are kinematically possiible and 
are dominant compared to 2 $\rightarrow$ 2 processes~\cite{Hattori:2016lqx}. 
The thermal relaxation time $\tau_{\text{eff}}$,
 can be defined from the relativistic transport 
equation in terms of distribution function in the 
strong magnetic field $\vec{B}=B\hat{z}$ as,
\begin{equation}\label{36}
\dfrac{d f_{q}}{d t}=C(f_{q})\equiv
-\dfrac{\delta f_q}{\tau_{\text{eff}}}.
\end{equation}
Here, $C(f_{q})$ represents the collision integral 
for the process under consideration.
For the $1\rightarrow 2$ processes $(  p+p^{'}\longrightarrow k$, where primed notation 
for antiquark), the 
thermal relaxation in the strong magnetic field can be defined as follows,
\begin{align}\label{36.1}
\tau_{\text{eff}}^{-1}(p_z)&=\sum_{l^{'}=0}^{\infty}\int_{-\infty}^{\infty}{\dfrac{dp^{'}_{z}}{2\pi}\int{\dfrac{d^3k}{(2\pi)^3}}
\dfrac{(2\pi)^2\delta(k_z-p_z-p^{'}_{z})}{2\omega_k2\omega_{l_p}2\omega_{l_{p^{'}}}}}\nonumber\\
&\times\mid M_{p+p^{'}\rightarrow k}\mid^2\dfrac{f^0_{q}(p^{'}_{z})( 1+f_{g}^{0}(k))}{(1-f_{q}^{0}(p_z))},
\end{align}
where the quasiquark distribution function is defined as,
\begin{equation}\label{1}
f^0_{q}=\dfrac{z_{q}\exp{(-\beta \sqrt{p_{z}^{2}+m_f^{2}+2l\mid q_feB\mid
})}}{1+ z_{q}\exp{( -\beta 
\sqrt{p_{z}^{2}+m_f^{2}+2l\mid q_feB\mid} )}},
\end{equation}
and the quasigluon
distribution function has the form,
\begin{equation}\label{2}
f^0_{g}=\dfrac{z_{g}\exp{(-\beta\mid\vec{k}\mid)}}
{1- z_{g}\exp{( -\beta\mid\vec{k}\mid} )}.
\end{equation}
Within the LLL approximation the momentum dependent thermal relaxation time takes the following form in the 
regime $p_{z^{'}}\sim 0$, as~\cite{Kurian:2018dbn,Hattori:2016cnt},
\begin{equation}\label{38}
{(\tau_{\text{eff}}^{-1})}_{l=0}=\dfrac{2\alpha_{\text{eff}}C_{F} m_f^2}{\omega_{q} 
(1-f_{q}^{0})}\dfrac{z_q}{(z_{q}+1)}
( 1+f_{g}^{0}(E_{p_z}))\ln{(T/m)}, 
\end{equation}
where $C_F$ is the Casimir factor of the processes
and $\alpha_{\text{eff}}$ is the effective coupling constant
defined from the Debye screening mass~\cite{Kurian:2018dbn}.

The Impact of the higher 
Landau levels on the 
matrix element and distribution function for the $1\rightarrow 2$ processes is explored 
in the very recent work~\cite{Fukushima:2017lvb}. Including these HLL effects, the 
thermal relaxation time of the $1\rightarrow 2$ processes has the following form,
\begin{align}\label{38.2}
\tau_{\text{eff}}^{-1}(p_z)&=\dfrac{1}{4\omega_{l_q}}\dfrac{1}{(1-f_{q}^{0}(p_z))}\sum_{l^{'}\geq l}^{\infty}\int_{-\infty}^{\infty}{\dfrac{dp^{'}_{z}}{2\pi}}
\dfrac{1}{2\omega_{{l^{'}}_{\bar{q}}}}X(l, l^{'},\xi)\nonumber\\
&\times f^0_{q}(p^{'}_{z})( 1+f_{g}^{0}(p^{'}_{z}+p_z)),
\end{align}
where $\xi$ is defined as,
\begin{equation}
\xi=\dfrac{(\omega_{l_q}+\omega_{{l^{'}}_{\bar{q}}})^2-(p_z+p^{'}_{z})^2}{2\mid q_feB\mid},
\end{equation}
and $X(l, l^{'},\xi)$ takes the form as follows,
\begin{align}\label{38.3}
X(l, l^{'},\xi)&=4\pi\alpha_{\text{eff}}N_cC_F\dfrac{l!}{l^{'}!}e^{-\xi}\xi^{l^{'}-l}\Bigg[\bigg(4m_f^2\nonumber\\
&-4\mid q_feB\mid 
(l+l^{'}-\xi)\dfrac{1}{\xi}(l+l^{'})\bigg)F(l, l^{'}, \xi)\nonumber\\
&+16\mid q_feB\mid l^{'}(l+l^{'})\dfrac{1}{\xi}L_{l}^{(l^{'}-l)}(\xi)L_{l-1}^{(l^{'}-l)}(\xi)\Bigg],
\end{align}
\begin{figure}[h]
\hspace{-1cm}
  \subfloat{\includegraphics[scale=0.32]{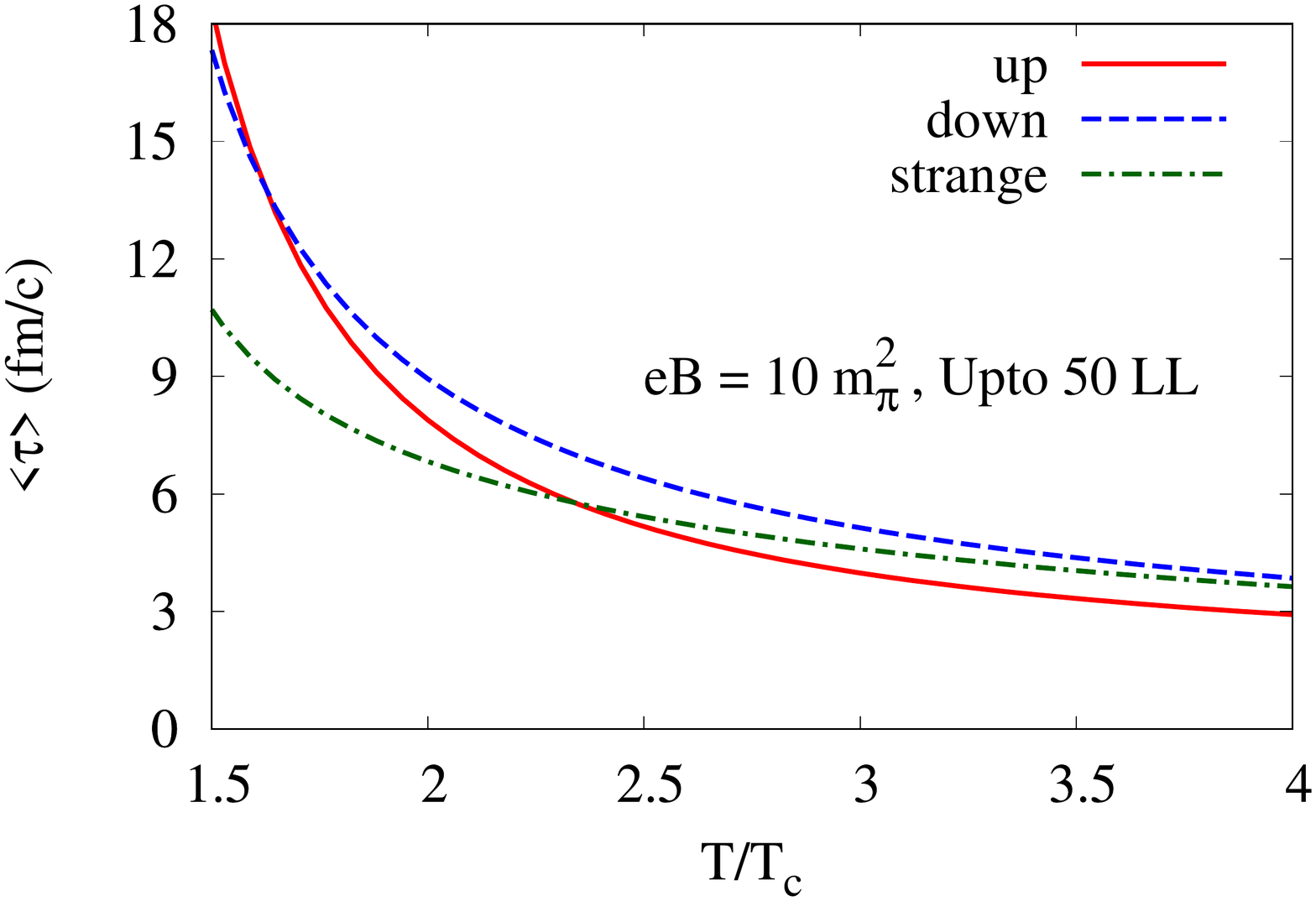}}
\caption{ The temperature dependence of thermal relaxation time for 
 quarks at $\mid eB\mid=10 m_{\pi}^2$.}
\label{f1}
\end{figure}
with $F(l, l^{'}, \xi)=[L_{l}^{(l^{'}-l)}(\xi)]^2+\dfrac{l^{'}}{l}[L_{l-1}^{(l^{'}-l)}(\xi)]^2$ for $l>0$ 
and $F(l, l^{'}, \xi)=1$ for the lowest Landau level. 
Here, $\alpha_{\text{eff}}$ is the effective 
coupling constant and is defined from the Debye 
screening masses of the QGP~\cite{Mitra:2017sjo, Bandyopadhyay:2016fyd, Ghosh:2018xhh,
Bonati:2017uvz, Singh:2017nfa}.  
\begin{figure}[h]
\hspace{-1cm}
  \subfloat{\includegraphics[scale=0.32]{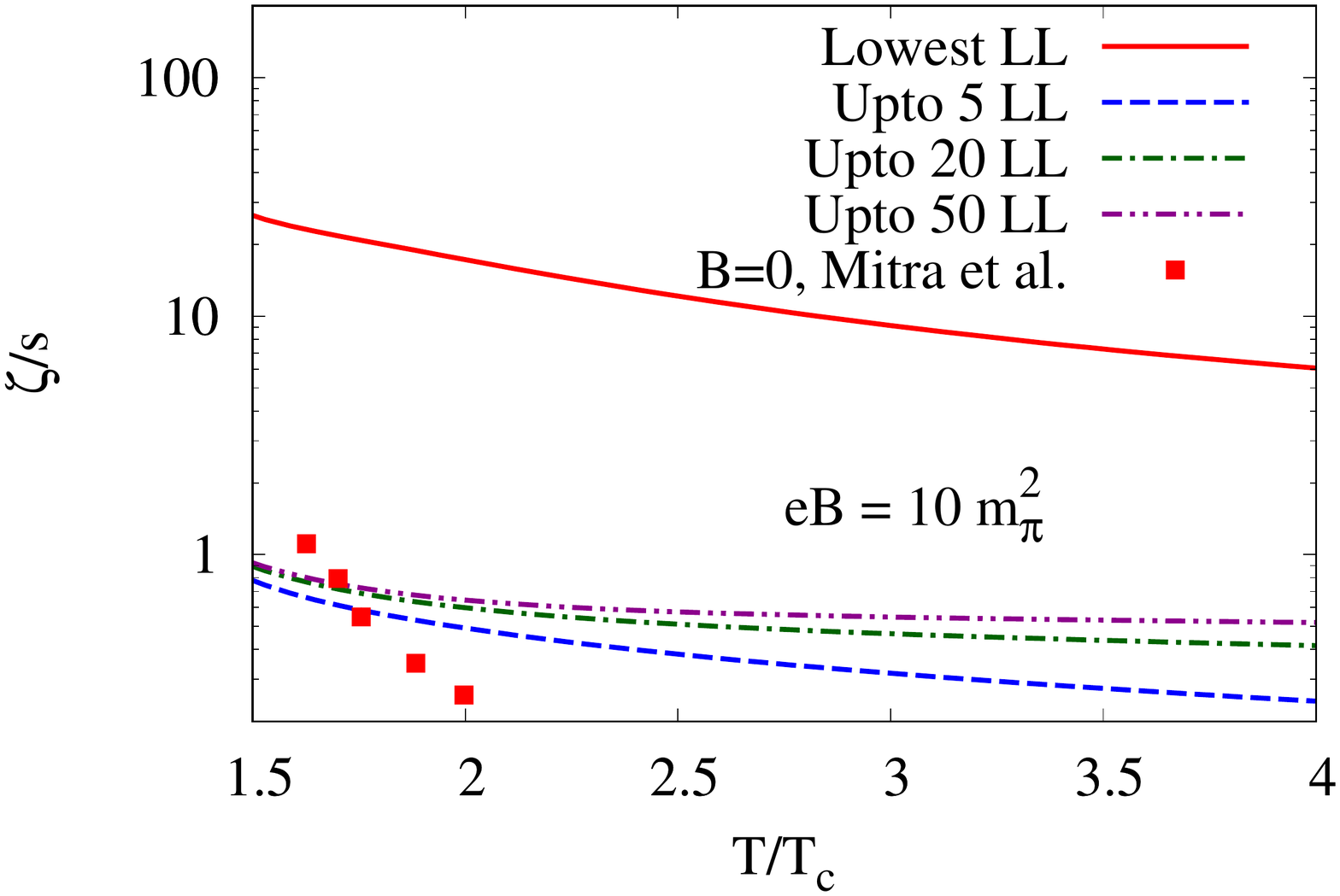}}
\caption{ The effects of HLLs on the  temperature behaviour of $\zeta/s$ at $\mid eB\mid=10 m_{\pi}^2$. 
Behaviour of $\zeta/s$ is comparing with the result at ${\bf{B}}=0$ of Mitra et al.~\cite{Mitra:2018akk}.}
\label{f2}
\end{figure}

\begin{figure}[h]
\hspace{-1cm}
  \subfloat{\includegraphics[scale=0.32]{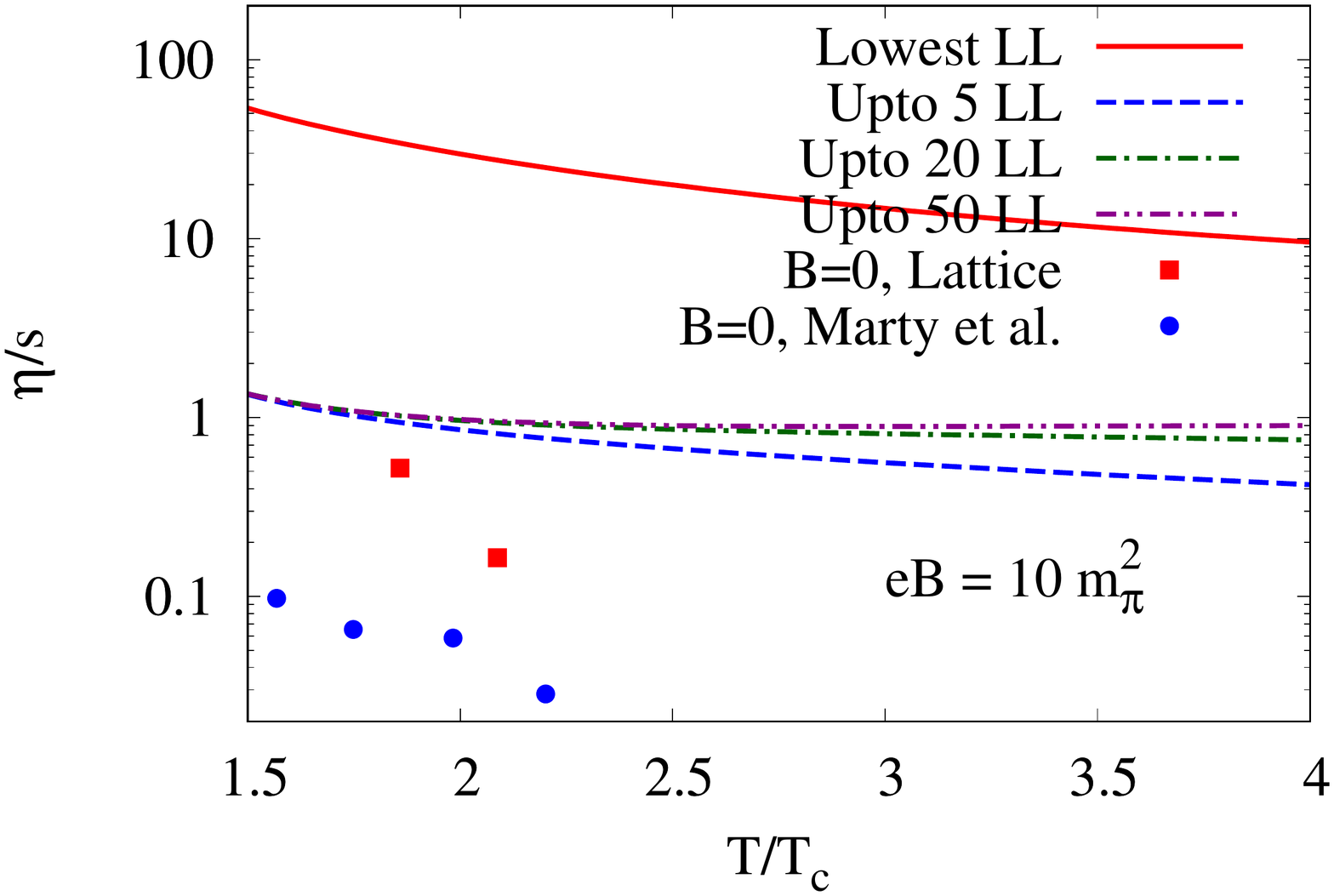}}
\caption{ The effects of HLLs on the  temperature behaviour of $\eta/s$ at $\mid eB\mid=10 m_{\pi}^2$. 
Lattice data~\cite{Nakamura:2004sy} and result of Marty et al.~\cite{Marty:2013ita} for $\eta/s$ are in the absence of magnetic field.}
\label{f3}
\end{figure}

Hot medium effects are entering through the quasiparton 
distribution function and the effective coupling.  The effective thermal relaxation time 
controls the behaviour 
of transport coefficients critically.
Note that in the limit $T^{2}\ll \mid q_feB\mid$, LLL approximation is valid so that
$X(l=0, l^{'}=0, \xi)\approx 16\pi{(\alpha_{\text{eff}})}m_f^2N_cC_F$, where 
$e^{-\xi}\approx 1$ in this regime. Hence, the 
thermal relaxation time as defined in the Eq.~(\ref{38.2}) can be reduced to the LLL result as defined in Eq.~(\ref{38}) 
in the limit $T^{2}\ll \mid q_feB\mid$. Following the parton distribution function within the EQPM framework, the 
thermal average of $\tau_{\text{eff}}$ can be defined as,
\begin{equation}
<\tau_{\text{eff}}>=\dfrac{\sum_{l=0}^{\infty}\int_{-\infty}^{\infty}{dp_z\tau_{\text{eff}}f^0_q}}{\sum_{l=0}^{\infty}\int_{-\infty}^{\infty}{dp_zf^0_q}}.
\end{equation}
Notably, the thermal average is taken merely to explore the temperature behaviour of $<\tau_{\text{eff}}>$ with the inclusion of  the effects of HLLs  and analysed in the next section. While computing the 
transport coefficients the momentum dependence of the relaxation time, $\tau_{eff}$ has been employed.

\section{Results and discussions}
Let us  initiate the discussion with the temperature behaviour of thermal 
relaxation time $\tau_{\text{eff}}$ of the quarks (up, down and strange quarks with 
masses $m_u=3$ MeV, $m_d=5$ MeV and $m_s=100$ MeV respectively) for the dominant 
$1\rightarrow 2$ processes in the presence of 
the strong magnetic field. Thermal relaxation time has been plotted 
as a function of $\frac{T}{T_c}$ for $\mid eB\mid=10 m^2_{\pi}$ considering up to 
50 LLs in the Fig.~\ref{f1}. The relaxation time exhibits the decreasing trend with increasing temperature. 
In the limit, $T^{2}\ll \mid q_feB\mid$, $\tau_{\text{eff}}$ defined in Eq.~(\ref{38.2}) reduced to the 
LLL result as described in~\cite{Kurian:2018dbn}. To encode the EoS effects in the thermal relaxation, 
the quasiparticle parton distribution functions are introduced along with the effective coupling constant. 
The thermal relaxation time act as the dynamical input for the transport processes.
\begin{figure*}
 \centering
 \subfloat{\includegraphics[scale=0.32]{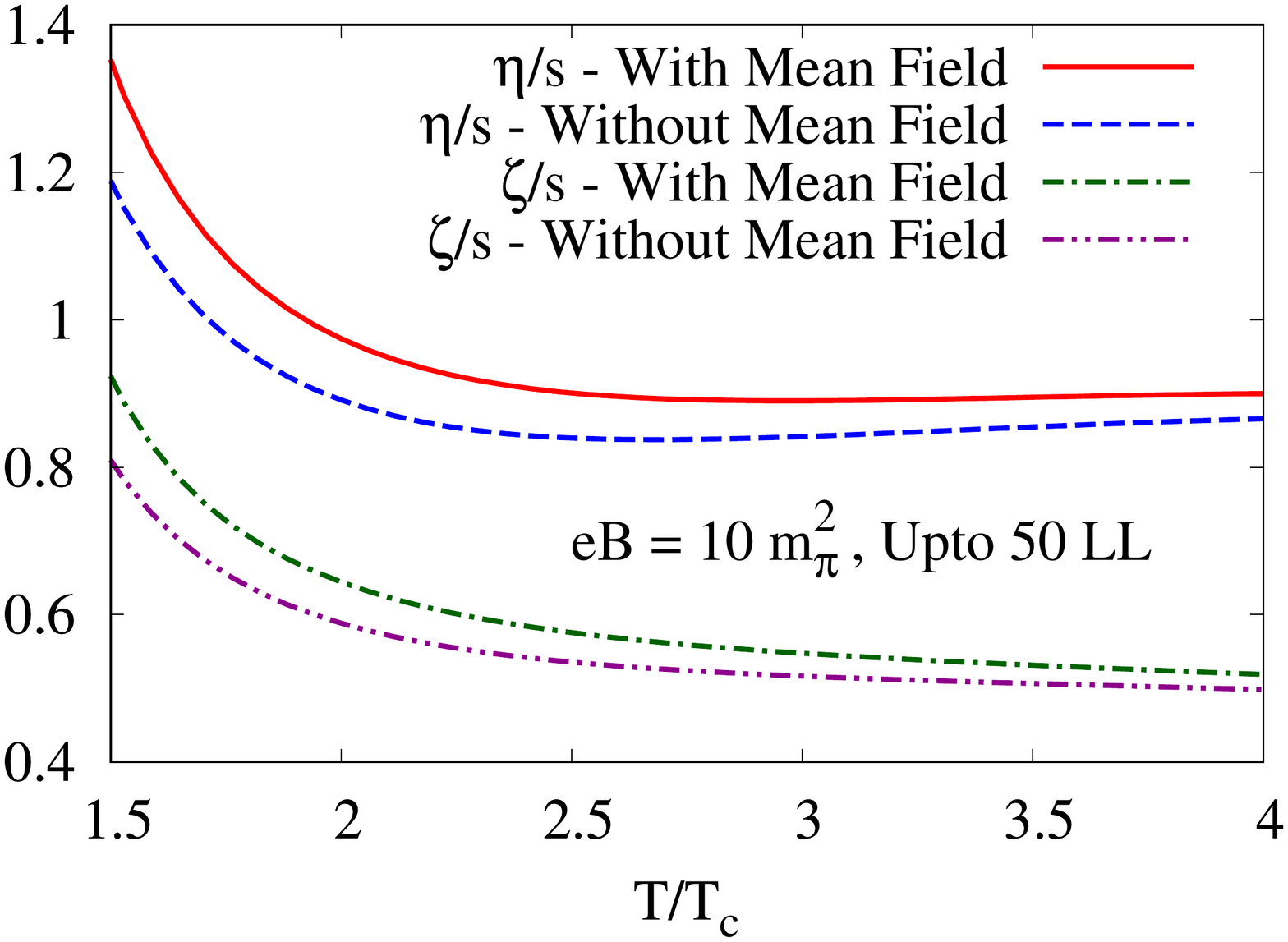}}
 \hspace{-1.5 cm}
 \subfloat{\includegraphics[scale=0.32]{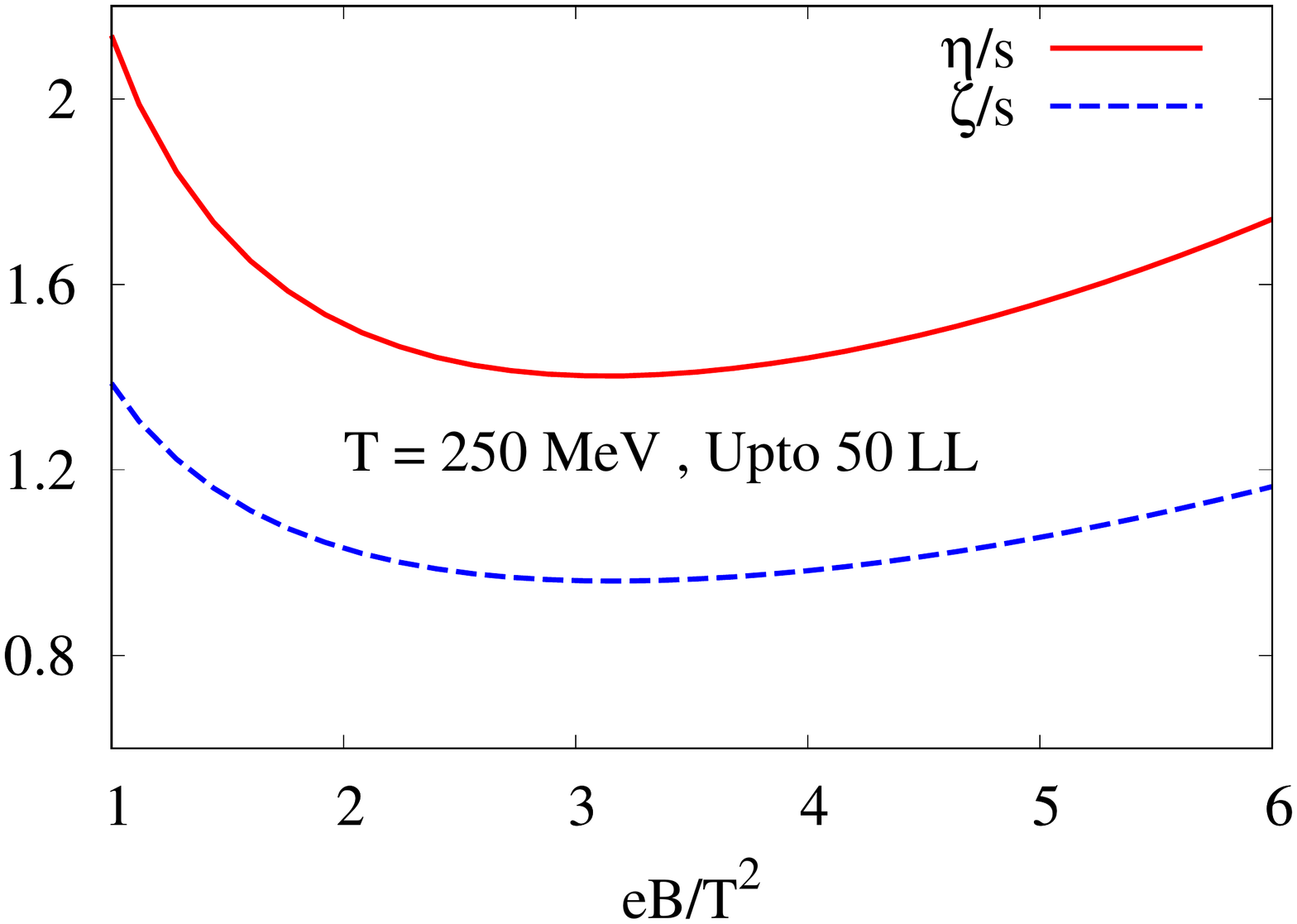}}
 \caption{ Temperature dependence of $\zeta/s$ 
 and $\eta/s$ with and without mean field correction at  $\mid eB\mid=10  m_{\pi}^2$ for 1$\rightarrow$ 2 processes  
 (left panel). Magnetic field dependence of $\zeta/s$ and $\eta/s$ (right panel). }
 \label{f4}
\end{figure*}      

Following the Eq.~(\ref{26}), the temperature dependence of bulk viscosity depends on the term 
$\frac{1}{\omega_p^2}(p_{z_k}^2-\omega_p^2c_s^2)^2$ 
and the relaxation time $\tau_{\text{eff}}$, where $c_s^2$ can be obtained 
from the QCD thermodynamics.
 The ratio of longitudinal bulk viscosity 
to entropy density for the $1\rightarrow2$ processes at $\mid eB\mid=10 m^2_{\pi}$ 
has been plotted as a function of $T/T_c$ in the Fig.~\ref{f2}. 
The temperature dependence of the $\zeta/s$ in the strong magnetic 
field indicates its rising behaviour near $T_c$. 
The behaviour of longitudinal shear viscosity for the 1$\rightarrow$ 2 processes with
 $T/T_c$ at $\mid eB\mid=10 m^2_{\pi}$ is shown in 
 Fig.~\ref{f3}. Since the driving force for the 
longitudinal shear viscosity is in the direction 
of the magnetic field, the Lorentz force does not interfere in the calculation. 
Quantitatively, $\eta/s$ with the HLL contributions remains 
within the same range of the lattice data~\cite{Nakamura:2004sy} 
 and NJL model result in~\cite{Marty:2013ita} at ${B}={0}$. This observation 
 is in line with the result that longitudinal conductivity with HLLs contributions 
 remains within the range of the lattice result at zero magnetic field~\cite{Fukushima:2017lvb}.
  For the numerical estimation of 
$\zeta/s$ and $\eta/s$, we truncate the Landau level sum at $l_{\text{max}}=50$. We observe that the HLL contributions are 
significant in the estimation of the viscous coefficients whereas the LLL approximation has an enhancement as 
$m_f$ tends to zero. Our observations on the effects of HLLs to the transport coefficients are qualitatively consistent with the results of 
the recent work of Fukushima and Hidaka~\cite{Fukushima:2017lvb}. 
\begin{figure}[h]
\hspace{-1cm}
  \subfloat{\includegraphics[scale=0.32]{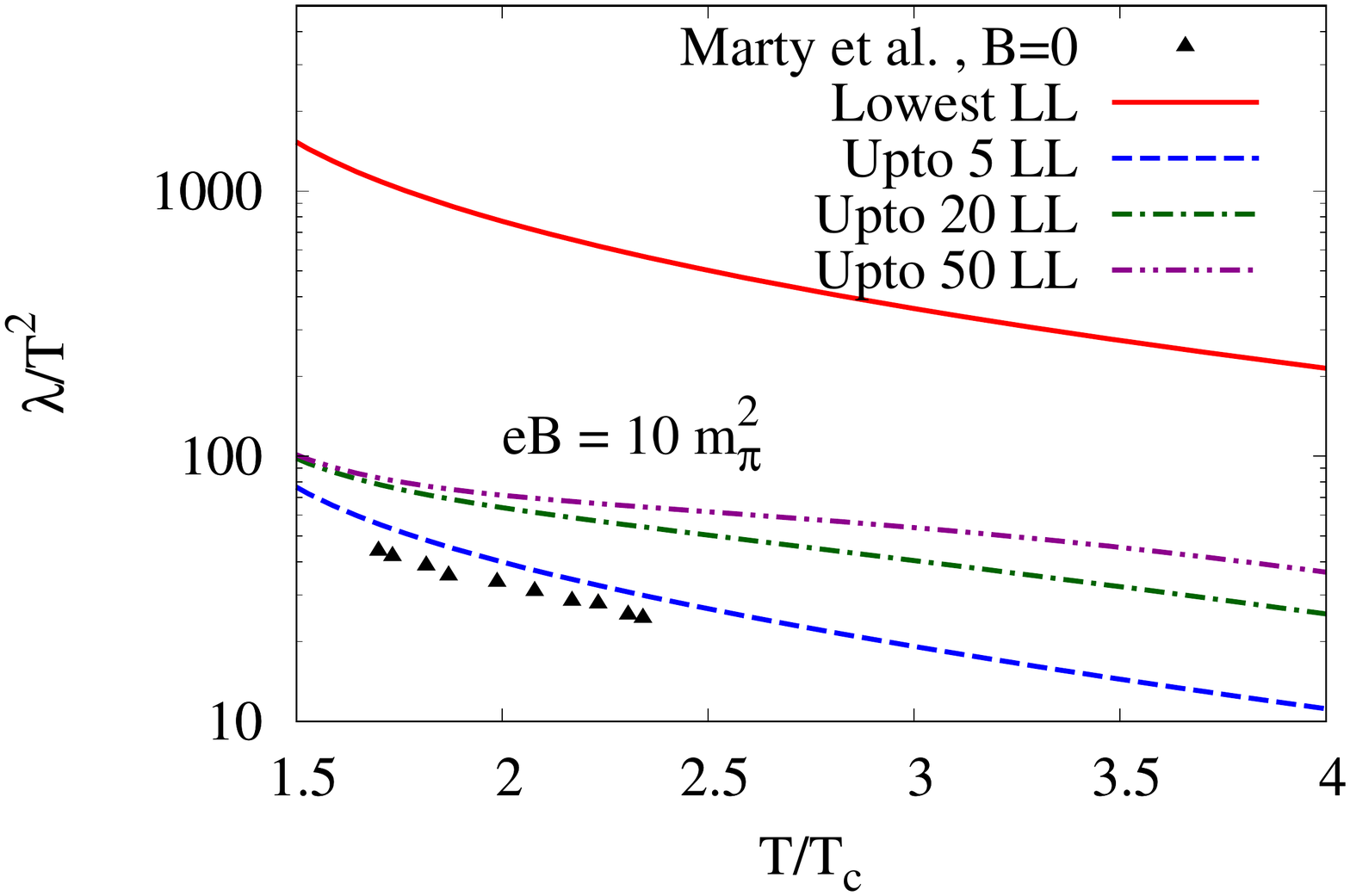}}
\caption{ Thermal conductivity as a function of $T/T_c$ at $\mid eB\mid=10  m_{\pi}^2$. Behaviour of $\lambda/T^2$ is 
comparing with the result at ${\bf{B}}=0$ of Marty et al.~\cite{Marty:2013ita}}
\label{f5}
\end{figure} 
 
The present analysis is done by employing the effective 
covariant kinetic theory using the Chapman-Enskog method including the effects of HLLs. 
The mean field force term which 
emerges from the effective theory indeed appears as the 
mean field corrections to the transport coefficients. 
The second term in the Eq.~(\ref{25}) and Eq.~(\ref{26}) describes the mean field 
contribution to the longitudinal shear viscosity and bulk viscosity in the 
strong magnetic field, respectively. The mean field term consists of the term $\delta\omega$ which is the 
temperature gradient of the effective fugacity $z_{g/q}$. 
The temperature behaviours of the viscous coefficients (bulk and shear viscosities) in the presence of 
strong magnetic field with and without the 
mean field corrections are shown in Fig.~\ref{f4} (left panel).
At higher temperature, the effects are negligible since the effective 
fugacity behaves as a slowly varying function of temperature there. 
Hence, the mean field corrections due to the quasiparticle 
excitations are significant at temperature region closer to  $T_c$.
The magnetic field dependence of the bulk viscosity and shear 
viscosity have been plotted in the Fig.~\ref{f4}(right panel).  In the 
strong magnetic field limit, the viscous coefficients  could be computed within   LLL approximation. The inclusion of HLLs reflects
the non-trivial (non-monotonic) magnetic field dependence of the transport coefficients. Similar non-monotonic 
structure in the magnetic field dependence of longitudinal 
conductivity with HLLs is described in~\cite{Fukushima:2017lvb}. The estimation of electric conductivity within our model while including the 
HLLs is beyond the scope of the present analysis and is a matter of future investigations.
 
    Mean field corrections to the thermal conductivity is 
  explicitly shown in Eq.~(\ref{30}) in which thermal 
  relaxation incorporates the microscopic interactions. 
  We depicted the temperature behaviour of $\lambda/T^2$ in Fig.~\ref{f5}. 
   The HLL effects of the transport coefficients are entering through the thermal relaxation time 
  and the quasiparticle distribution function. These effects are significant in the estimation of 
  transport coefficients in the presence of a magnetic field. The temperature behaviour of the 
  dimensionless quantity $\lambda/T^2$  
  in the absence of the magnetic field is well investigated~\cite{Mitra:2017sjo, Marty:2013ita} and is in the 
  order of $100-25$ within the temperature range $(1-4)\frac{T}{T_c}$, which is quantitatively consistent with our result.

\section{Conclusion and Outlook}
In conclusion, we have computed the temperature behaviour 
of the transport parameters such as longitudinal viscous 
coefficients (shear and bulk viscosities) 
and thermal conductivity for the $1\rightarrow2$ processes in the strong magnetic field 
background while including the effects of HLLs.
 Thermal relaxation time is computed in the strong magnetic field incorporating the 
HLL contributions. Setting up an effective covariant 
kinetic theory within EQPM in the strong magnetic 
field induces mean field contributions to the 
transport coefficients. We employed the Chapman-Enskog 
method in the effective kinetic theory for the 
computation of transport coefficients. The 
transport coefficients that have been estimated 
are influenced by the thermal medium and magnetic field. Hot QCD 
effects are incorporated through the quasiparton 
degrees of freedom along with effective coupling 
and the medium effects are found to be negligible 
at very high temperature.   We focused on the weakly coupled regime of the perturbative 
QCD within the limit $gT
\ll \sqrt{\mid q_feB\mid}$ in which higher Landau level (HLL) contributions are significant.
Notably, the inclusion of HLL contributions are essential to explain the transport processes
at high temperature in the presence of the magnetic field.
Furthermore, effects of the mean field term are seen to be quite significant as fas as the temperature behavior of the  above mentioned transport coefficients is concerned (for the temperatures which are
not very far away from $T_c$). 

An immediate future extension of the work is to 
investigate the aspects of non-linear electromagnetic 
responses of the hot QGP with the mean field 
contribution along with the effective description 
of magnetohydrodynamic waves in the hot QGP medium. 
In addition, the estimation of all
transport coefficients from covariant kinetic 
theory within the effective fugacity 
quasiparticle model using more realistic 
collision integral, for example, BGK 
(Bhatnagar, Gross and Krook) collision term, 
in the strong magnetic field 
would be another direction to work.

\section*{acknowledgments}
V.C. would like to acknowledge Science 
and Engineering Research
Board (SERB), Govt. of India for the Early
Career Research Award (ECRA/2016) and Department of Science and
Technology (DST), Govt. of India for INSPIRE-Faculty
Fellowship (IFA-13/PH-55).  S.G. would to like acknowledge the Indian Institute of Technology 
Gandhinagar for the postdoctoral fellowship.
S.M. would like to acknowledge 
SERB-INDO US forum to conduct the 
Postdoctoral research in USA.  We record our gratitude to the people of India 
for their generous support 
for the research in basic sciences.

{}

\end{document}